\documentclass[letterpaper,twocolumn,prl,aps,showpacs,floatfix,superscriptaddress,nofootinbib]{revtex4-1}
\usepackage[latin1]{inputenc}
\usepackage{bm}
\usepackage{multirow}
\usepackage{amssymb}
\usepackage{amsbsy}
\usepackage{amsmath}
\usepackage{graphicx}
\usepackage{epsfig}
\usepackage{placeins}
\usepackage{ulem}
\usepackage[usenames,dvipsnames]{color}
\usepackage[colorlinks,linkcolor=blue,citecolor=blue,urlcolor=blue]{hyperref}
\makeatletter

\begin{document} 
\title{Charge diffusion in the one-dimensional Hubbard model}

\author{R. Steinigeweg}
\email{rsteinig@uos.de}
\affiliation{Department of Physics, University of Osnabr\"uck, D-49069 Osnabr\"uck, Germany}

\author{F. Jin}
\affiliation{Institute for Advanced Simulation, J\"ulich Supercomputing Centre, Forschungszentrum J\"ulich, D-52425 J\"ulich, Germany}

\author{H. De Raedt}
\affiliation{Zernike Institute for Advanced Materials, University of Groningen, NL-9747AG Groningen, The Netherlands}

\author{K. Michielsen}
\affiliation{Institute for Advanced Simulation, J\"ulich Supercomputing Centre, Forschungszentrum J\"ulich, D-52425 J\"ulich, Germany}
\affiliation{RWTH Aachen University, D-52056 Aachen, Germany}

\author{J. Gemmer}
\email{jgemmer@uos.de}
\affiliation{Department of Physics, University of Osnabr\"uck, D-49069 Osnabr\"uck, Germany}

\begin{abstract}
We study the real-time and real-space dynamics of charge in the one-dimensional
Hubbard model in the limit of high temperatures. To this end, we prepare pure
initial states with sharply peaked density profiles and calculate the time
evolution of these nonequilibrium states, by using numerical forward-propagation
approaches to chains as long as $20$ sites. For a class of typical states,
we find excellent agreement with linear-response theory and unveil the existence
of remarkably clean charge diffusion in the regime of strong particle-particle
interactions. Moreover, we demonstrate that this diffusive behavior does not
depend on certain details of our initial conditions, i.e., it occurs for five
different realizations with random and nonrandom internal degrees of freedom,
single and double occupation of the central site, and displacement of spin-up
and spin-down particles.
\end{abstract}


\maketitle

{\it Introduction.} Static properties of integrable quantum many-body systems
are well understood \cite{johnston2000}. In contrast, dynamical questions in
these systems continue to be a major challenge in many areas of modern physics
and range from fundamental questions in statistical physics to applied questions
for a specific class of materials. On the one hand, integrable systems feature
a macroscopic number of (quasi-)local conservation laws \cite{zotos1997,
prosen2011, prosen2013} and any overlap with these conserved quantities leads
to the breakdown of conventional equilibration and thermalization
\cite{rigol2007, ilievski2015}. On the other hand, such overlap is not warranted
for all possible initial states, observables, or model parameters and
integrability does not rigorously exclude the existence of thermodynamic
relaxation such as exponential decay or diffusive transport. This type of
relaxation, however, is often traced back to chaos \cite{mejiamonasterio2005,
dalessio2016}, being absent in integrable systems.

In this context, two important and extensively studied examples are (i) the
one-dimensional XXZ spin-$1/2$ model and (ii) the (Fermi-)Hubbard chain. As
typical for integrable systems, the energy current is (i) strictly or (ii)
at least partially conserved \cite{zotos1997} such that energy flow is
ballistic at any finite temperature \cite{kluemper2002, langer2011,
karrasch2014-1, karrasch2016-1}, as signaled by a nonzero Drude weight within
linear-response theory. However, a much richer dynamical phase diagram develops
for other transport quantities. In case (i) of the XXZ spin-$1/2$ chain, the
spin current is not strictly conserved. While the partial conservation of this
current and a nonzero Drude have been proven analytically below the isotropic
point \cite{zotos1999, benz2005, prosen2011, prosen2013}, strong numerical
evidence for a vanishing Drude weight and nonballistic dynamics has been
provided above this point \cite{heidrichmeisner2003, herbrych2011, karrasch2012,
steinigeweg2014-1, steinigeweg2015}. In fact, for the latter regime, clear
signatures of diffusion have been reported in various works \cite{znidaric2011,
steinigeweg2011, karrasch2014-1, steinigeweg2016}. In case (ii) of the Hubbard
chain, the situation appears to be similar for charge transport. Even though
clarifying the existence of a nonzero Drude weight has turned out to be hard
task analytically \cite{fujimoto1997, kirchner1999, peres2000, carmelo2013},
numerical studies point to a vanishing Drude weight for strong particle-particle
interactions \cite{prelovsek2004, karrasch2014-2, jin2015}. While signatures of
diffusion have been observed for such interactions also \cite{jin2015,
karrasch2016-2, prosen2012, prosen2014}, a direct detection of the
characteristic Gaussian broadening is lacking.

The intention of our Letter is to clarify the existence of charge diffusion in
the Hubbard chain. For this purpose, we study the nonequilibrium dynamics as
resulting for a convenient class of initial states. These initial states are pure
and realize density profiles where a peak with the maximum amplitude possible is
located in the center of the chain and lies on top of a homogeneous many-body
background. First, we focus on a subclass with random internal degrees of
freedom and rely on the well-known concept of typicality \cite{gemmer2003,
goldstein2006, popescu2006, reimann2007, bartsch2009, bartsch2011, sugiura2012,
sugiura2013, elsayed2013, hams2000, iitaka2003, iitaka2004, white2009,
monnai2014} to obtain the real-time broadening of density profiles in the
linear-response regime. In this regime, our large-scale numerical simulations
for chains as long as $20$ sites allow us to unveil the existence of remarkably
clean charge diffusion, as a key result of our Letter. Finally, we extend our
analysis to initial states without any randomness and show that the dynamical
behavior is stable against varying details of the initial conditions. This
stability is another central result of our work and reveals that exactly the
same charge diffusion emerges in a far-from-equilibrium setup. These findings
clearly demonstrate that thermodynamic relaxation can occur in integrable
systems.

{\it Model and Observables.} In one spatial dimension and with periodic boundary
conditions, the Hamiltonian of the Hubbard model reads $H = \sum_{r=1}^L h_r$ ,
\begin{equation}
h_r = -t_\text{h} \sum_{s=\downarrow, \uparrow} (a_{r,s}^\dagger a_{r+1,s} +
\text{H.c.}) + U (n_{r,\downarrow} - \frac{1}{2})(n_{r,\uparrow} - \frac{1}{2})
\, , \label{H}
\end{equation}
where the operator $a_{r,s}^\dagger$ ($a_{r,s}$) creates (annihilates) at site
$r$ a fermion with spin $s$, $t_\text{h}$ is the hopping matrix element, and $L$
is the number of sites. The operator $n_{r,s} = a_{r,s}^\dagger a_{r,s}$ is the
local occupation number and $U$ is the strength of the on-site interaction. For
all values, this model is integrable and the total particle numbers $N_s =
\sum_r n_{r,s}$ and $N = N_\downarrow + N_\uparrow$ are strictly conserved
quantities. We do not restrict ourselves to a particular particle sector, i.e.,
we study the case $\langle N \rangle = L$ \cite{SM}. It is worth mentioning
that, via the Jordan-Wigner transformation, this model can be mapped onto a
spin-$1/2$ two-leg ladder of length $L$, with XY exchange in the legs and
Ising exchange in the rungs \cite{prosen2012}. In fact, this spin model is used
in our numerical simulations.

We are interested in the real-time dynamics of the local occupation numbers
$n_{r,s}$ and investigate the expectation values $p_{r,s}(t) = \text{tr}[n_{r,s}
\, \rho(t)]$ for the density matrix $\rho(t)$ at time $t$. (It is important to
note that $t \neq t_\text{h}$.) Doing so, we can follow the broadening of
nonequilibrium density profiles, as realized by the preparation of a proper
initial state $\rho(0)$.

{\it Initial States.} In this Letter, we prepare pure initial states $\rho(0)
= | \psi(0) \rangle \langle \psi(0) |$. To specify our $| \psi(0) \rangle$,
it is convenient to consider the common eigenbasis of all $n_{r,s}$. Let $|
\varphi_k \rangle$ be this basis. Then, our initial states read
\begin{equation}
| \psi(0) \rangle \propto n_{L/2,\uparrow} \, | \phi \rangle \, , \quad  | \phi
\rangle = \sum_{k=1}^{4^L} c_k \, | \psi_k \rangle \, , \label{single}
\end{equation}
where $c_k$ are complex and yet arbitrary coefficients. Since $n_{L/2,\uparrow}$
projects only onto states with a spin-up particle in the middle of the chain,
$p_{L/2,\uparrow}(0) = 1$ has the maximum value possible.

In this Letter, we focus on two particular choices for the coefficients $c_k$.
First, we choose all $c_k$ to be the same number. Second, we choose the $c_k$ at
random according to the unitary invariant Haar measure \cite{bartsch2009}. This
choice means in practice that real and imaginary part of the $c_k$ are
independently drawn from a Gaussian distribution with zero mean. For both, the
random and equal choice of the $c_k$, all $p_{r\neq L/2, s\neq \uparrow}(0) =
p_\text{eq.} = 1/2$ take on their equilibrium value, and still $p_{L/2,
\uparrow}(0) = 1$. Therefore, the initial density profile features a central peak
on top of a homogeneous many-particle background.

A very similar form for the density profile also results for the state
\begin{equation}
| \psi_\text{double}(0) \rangle \propto n_{L/2,\downarrow} \, | \psi(0) \rangle
\, . \label{double}
\end{equation}
Then, due to the additional projection, $p_{L/2,\downarrow}(0) = 1$ also for
equal and random $c_k$. Therefore, the density profiles $p_{r,\uparrow}(t) =
p_{r,\downarrow}(t)$ are identical for $t = 0$ and all later times $t > 0$ as
well, see \cite{SM} for initial displacement.

\begin{figure}[t]
\includegraphics[width=0.95\columnwidth]{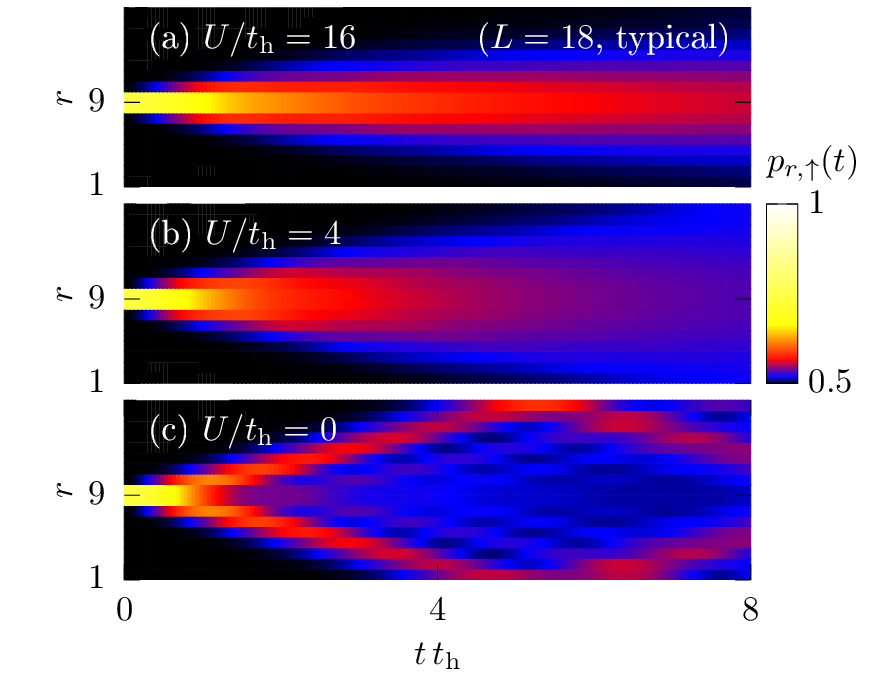}
\caption{(Color online) Time-space density plot of the spin-up occupation
numbers $p_{r,\uparrow}(t)$ for a {\it typical} initial state $|\psi(0) \rangle$
(where all spin-down occupation numbers $p_{r,\downarrow}(0) = p_\text{eq.}$) in
the one-dimensional Hubbard model with $L=18$ sites and different interaction
strengths: (a) $U/t_\text{h} = 16$, (b) $U/t_\text{h}= 4$, (c) $U/t_\text{h} =
0$. While the broadening in (a) points to charge diffusion, the broadening in
(c) is clearly ballistic.}
\label{Fig1}
\end{figure}

All initial states introduced have to be considered as far-from-equilibrium
states: They are not only pure but also have maximum $p_{L/2,\uparrow}(0)
= 1$. Remarkably, however, the dynamics of $| \psi(0) \rangle$ in Eq.\
(\ref{single}) with random $c_k$ can be connected to the linear-response
Kubo formula, since the underlying $| \phi \rangle$ is a so-called typical
state \cite{steinigeweg2016}, see also Refs.\ \cite{gemmer2003, goldstein2006,
popescu2006, reimann2007, bartsch2009, bartsch2011, sugiura2012, sugiura2013,
elsayed2013, hams2000, iitaka2003, iitaka2004, white2009, monnai2014} for the
concept of typicality. Exploiting this typicality allows one to derive the
relation \cite{SM}
\begin{equation}
p_{r,\uparrow}(t) -  p_\text{eq.} = 2 \, \langle (n_{L/2,\uparrow} -
p_\text{eq.})(n_{r,\uparrow}(t) - p_\text{eq.}) \rangle \, , \label{cor}
\end{equation}
where $\langle \bullet \rangle = \text{tr}[\bullet]/ 4^L$ is the thermodynamic
average at formally infinite temperature. Thus, for a typical state, the
nonequilibrium expectation value is directly related to an equilibrium
correlation function. This fact enables a connection to the Kubo formula
via the variance
\begin{equation}
\sigma(t)^2 = \sum_{r=1}^L r^2 \, \delta p_{r,\uparrow}(t) - \Big[ \sum_{r=1}^L
r \, \delta p_{r,\uparrow}(t) \Big ]^2 \, ,\label{variance}
\end{equation}
where $\delta p_{r,\uparrow}(t) = 2(p_{r,\uparrow}(t) - p_\text{eq.})$ excludes
the equilibrium background and is normalized to $\sum_r \delta p_{r,\uparrow}(t)
= 1$. As shown in Ref.\ \cite{steinigeweg2009}, the time derivative of this
variance satisfies
\begin{equation}
\frac{\text{d}}{\text{d}t} \sigma(t)^2 = 2 \, D(t) \, , \quad D(t) =
\frac{4}{L} \int_0^t \text{d}t' \, \langle j_\uparrow(t') j_\uparrow \rangle
\, , \label{D}
\end{equation}
where $j_\uparrow = - t_\text{h} \sum_r (\imath a_{r,\uparrow}^\dagger a_{r+1,
\uparrow} + \text{H.c.})$ is the total current of the spin-up particles and the
quantity $D(t)$ plays the role of a time-dependent diffusion coefficient. For
$U = 0$, [$j_\uparrow, H] = 0$ necessarily leads to ballistic scaling $D(t)
\propto t$ and $\sigma(t) \propto t$. For large $U \gg t_\text{h}$, signatures
of diffusive scaling $D(t) = \text{const.}$ and $\sigma(t) \propto \sqrt{t}$
have been reported in Refs.\ \cite{jin2015, karrasch2016-2}. So far, however,
a systematic analysis beyond the mere width of the density profile is lacking
and the central issue of our Letter.

\begin{figure}[t]
\includegraphics[width=0.95\columnwidth]{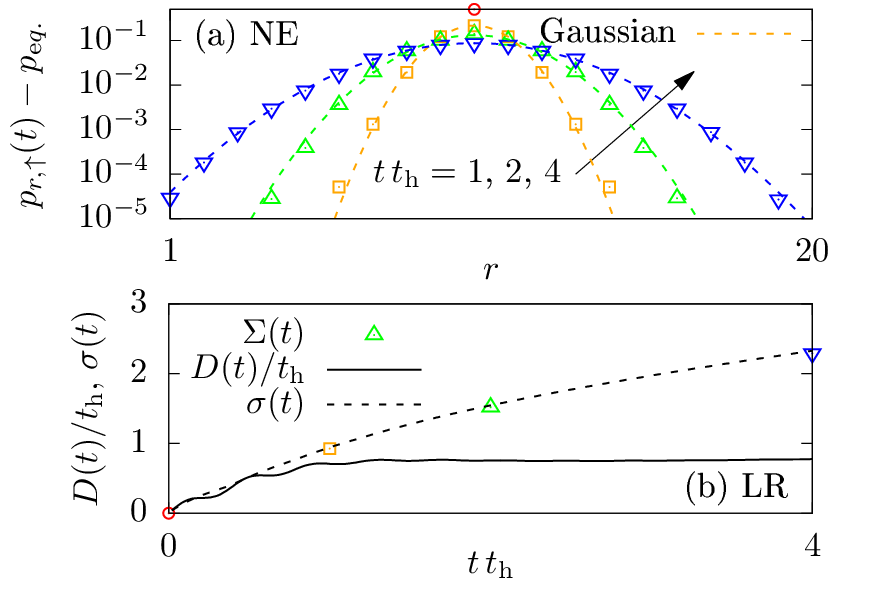}
\caption{(Color online) (a) Density profile $p_{r,\uparrow}(t)$ as a function of
site $r$ at various times $t \, t_\text{h} = 0, 1, 2, 4$ for a Hubbard chain of
length $L=20$ and with a strong interaction $U/t_h = 16$, shown in a semi-log
plot (symbols). (The initial state $| \psi(0) \rangle$ is the same as the one
in Fig.\ \ref{Fig1}.) The data can be described by Gaussian fits over several
orders of magnitude (curves). (b) Linear-response result for the time evolution
of the diffusion coefficient $D(t)$ and profile width $\sigma(t)$, as obtained
in Ref.\ \cite{jin2015} for length $L = 18 \sim 20$ and the same interaction
$U/t_\text{h} = 16$ (curves). The standard deviation $\Sigma(t)$, as resulting
from the Gaussian fits in (a), is indicated for comparison (symbols).}
\label{Fig2}
\end{figure}

{\it Numerical Technique and Results.} From a numerical point of view, the
Hubbard chain is challenging since the Hilbert-space dimension $\text{dim} =
4^L$ grows rapidly with $L$, e.g., much faster than the also exponential
increase $\text{dim} = 2^L$ in case of a spin-$1/2$ chain. As a consequence,
exact diagonalization of the Hamiltonian is only feasible for a few lattice
sites and a real-space experiment like the one done in our Letter would not
be reasonable. Hence, we proceed differently and profit from the fact that
we only need to deal with pure states. The time evolution of these states
can be obtained by forward-propagation methods such as fourth-order
Runge-Kutta \cite{steinigeweg2014-1, steinigeweg2015, elsayed2013} or more
sophisticated schemes such as Trotter decompositions or Chebyshev polynomials
\cite{steinigeweg2014-2, jin2015, weisse2006}. We apply a second-order Trotter
formula with a time step $\delta t \, t_\text{h} = 0.05$, sufficient to reach
very good agreement with Chebyshev-polynomial algorithms. A massively
parallelized implementation of this formula allows us to treat Hubbard chains
as long as $L=20$ sites. For this system size and a maximum time $t \,
t_\text{h} = 8$, the simulation takes about $9$ hours when using $262,144$
double-thread cores. Therefore, apart from the $L=20$ data depicted in Fig.\
\ref{Fig2}, we focus on $L = 18$ to reduce computational costs at least a bit.

Now, we turn to our numerical results. We start with the initial state $|
\psi(0) \rangle$ in Eq.\ (\ref{single}) with a random choice of the coefficients
$c_k$, i.e., a typical state. It is important to note that we consider a single
realization of the $c_k$ and do not perform any kind of averaging. Still, we
allow the $c_k$ to be different for each simulation. In Figs.\ \ref{Fig1}
(a)-(c) we depict our results for the spin-up occupation numbers $p_{r,
\uparrow}$(t) for a Hubbard chain of length $L=18$ and with different
interactions $U/t_\text{h} = 16$, $4$, $0$, in a 2D time-space density plot.
Several comments are in order. First, for the noninteracting case $U = 0$ in
Fig.\ \ref{Fig1} (c), the real-time broadening of $p_{r,\uparrow}(t)$ is clearly
linear and, as discussed above, has to occur due to the strict conservation of
the particle current. The pronounced jets visible are typical for free-particle
cases \cite{karrasch2014-1, steinigeweg2016} and propagate fast without any
scattering until they eventually hit the boundary of the chain at short times
$t \, t_\text{h} \sim 4$. Second, for the interacting cases $U/t_\text{h} = 16$,
$4$ in Figs.\ \ref{Fig1} (a), (b), these jets and the linear broadening as well
disappear, i.e., the dynamics is not ballistic. Note that, in contrast, the flow
of energy is ballistic for arbitrary $U$ \cite{SM}. Third, the broadening is the
slower the larger $U$ because scattering becomes stronger as $U$ increases. In
particular, for the largest $U/t_\text{h} = 16$ in Fig.\ \ref{Fig1} (a) and the
maximum time $t \, t_\text{h} = 8$ calculated, the overall width of $p_{r,
\uparrow}(t)$ is still smaller than the chain length. Therefore, we can exclude
trivial finite-size effects for such times \cite{SM}.

\begin{figure}[t]
\includegraphics[width=0.95\columnwidth]{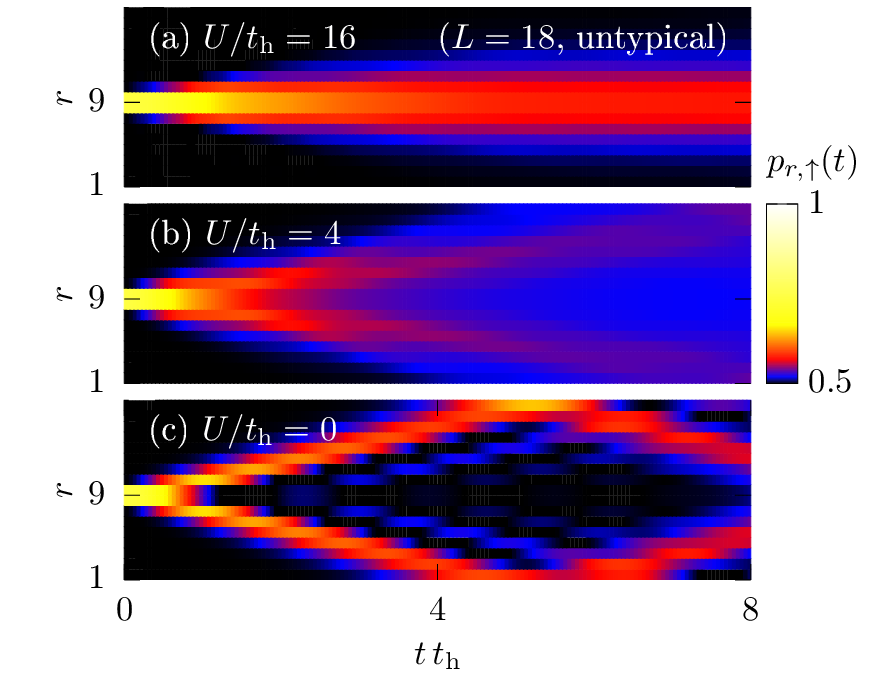}
\caption{(Color online) Time-space density plot of the spin-up occupation
numbers $p_{r,\uparrow}(t)$ for the same model parameters as in Fig.\ \ref{Fig1}
but for another and {\it untypical} initial state $| \psi(0) \rangle$ (where all
spin-down occupation numbers $p_{r,\downarrow}(0) = p_\text{eq.}$ once again).
Compared to Fig.\ \ref{Fig1}, jet-like behavior is enhanced in (c) while no
significant difference is visible in (a).}
\label{Fig3}
\end{figure}

To gain insight into the dynamics at large $U/t_\text{h}=16$, we show in Fig.\
\ref{Fig2} (a) the site dependence of the profile $p_{r,\uparrow}(t)$ for various
times $t \, t_\text{h} = 0$, $1$, $2$, $4$, and for an even larger system size
$L=20$. We do so by subtracting from $p_{r,\uparrow}(t)$ the equilibrium value
$p_\text{eq.}$ and using a semi-log plot, to visualize also the outer tails of
the profile. It is intriguing to see that, for all times $t$ depicted, the
profiles can be described very well by Gaussians
\begin{equation}
p_r(t) - p_\text{eq.} = \frac{1}{2} \, \frac{1}{\sqrt{2 \pi} \, \Sigma(t)}
\, \exp \! \left[-\frac{(r - L/2)^2}{2 \, \Sigma(t)^2} \right ] \, ,
\end{equation}
where the standard deviation $\Sigma(t)$ occurs as the only free parameter
and is adjusted by fitting. The excellent fits over several orders of magnitude
are a central result of our Letter and already provide strong evidence
for the existence of diffusion. Still, however, $\Sigma(t)$ needs to scale as
$\Sigma(t) \propto \sqrt{t}$. 

For a final conclusion, we show in Fig.\ \ref{Fig2} (b) the time dependence
of $\Sigma(t)$ and compare to the linear-response $\sigma(t)$ in Eq.\ (\ref{D}),
as resulting from the $D(t)$ calculated in Ref.\ \cite{jin2015} for the same
interaction $U/t_\text{h}=16$ and length $L=18$. While the perfect agreement
illustrates the high accuracy of the typicality relation, this agreement implies
that the linear-response result $D(t) = \text{const.}$ \cite{jin2015,
karrasch2016-2} also holds for our nonequilibrium dynamics. Thus, diffusion
clearly exists. Note that the same conclusion can be drawn for smaller
interactions $U/t_\text{h} = 8$ also \cite{SM}, where finite-size effects are
still negligibly small. For $U/t_\text{h} \ll 8$, however, significant
finite-size effects are known to occur \cite{jin2015} and a reliable conclusion
on the thermodynamic limit $L \to \infty$ is impossible on the basis of $L
\sim 20$.

\begin{figure}[t]
\includegraphics[width=0.95\columnwidth]{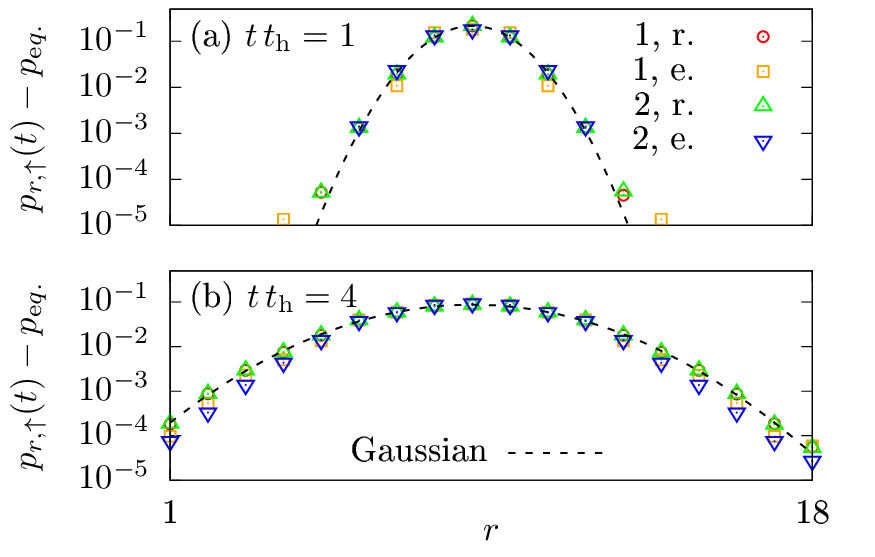}
\caption{(Color online) Density profile $p_{r,\uparrow}(t)$ versus site $r$ for
fixed times (a) $t \, t_\text{h} = 1$, (b) $t \, t_\text{h} = 4$ for various
initial states, i.e., $| \psi(0) \rangle$ (1) and $| \psi_\text{double}(0)
\rangle$ (2) with random (r.) and equal (e.) coefficients in the underlying
superposition. As a guide to the eyes, Gaussian fits are indicated (for 1,
r.). While data for random initial states are practically indistinguishable,
data for nonrandom initial states differ only very little.}
\label{Fig4}
\end{figure}

Next, we intend to shed light on the role of the specific initial-state
realization, in particular on the influence of randomness. Therefore, in a first
step, we investigate the initial state $| \psi(0) \rangle$ in Eq.\ (\ref{single})
again but now with equal coefficients $c_k$. Recall that, while this nonrandom
state has exactly the same initial density profile, the typicality relation does
not need to hold any further. For this state, we repeat the calculation in Fig.\
\ref{Fig1} for $L = 18$ sites and different interactions $U/t_\text{h} = 16$,
$4$, $0$ and summarize the corresponding results in Fig.\ \ref{Fig3}. In
comparison to Fig.\ \ref{Fig1}, jet-like behavior for $U = 0$ is enhanced in
Fig.\ \ref{Fig3} (c) and emerges now for $U/t_\text{h} = 4$ in Fig.\ \ref{Fig3}
(b) in addition. The same observation has been made for the XXZ spin-$1/2$ chain
below the isotropic point \cite{steinigeweg2016}. Remarkably, however, the
diffusive behavior for $U/t_\text{h} = 16$ in Fig.\ \ref{Fig3} (a) turns out
to be unaltered. In fact, this observation is different to the one found for
the XXZ spin-$1/2$ chain above the isotropic point \cite{steinigeweg2016},
where the impact of nonrandomness is strong.

The above finding suggests that charge diffusion for strong interactions $U$ is
stable against varying details of the initial condition. To substantiate this
suggestion, we finally extend our analysis to the initial state $|
\psi_\text{double}(0) \rangle$ in Eq.\ (\ref{double}) and study both, random and
equal coefficients $c_k$. To repeat, these states have the same initial density
profile $p_{r,\uparrow}(0)$ but $p_{r,\downarrow}(0) = p_{r,\uparrow}(0)$ now.
For system size $L=18$, interaction $U/t_\text{h} = 16$, and two different times
$t \, t_\text{h} = 1$, $4$, we compare in Fig.\ \ref{Fig4} the distribution
$p_{r,\uparrow}(t)$ for these states with the one for the others. Apparently,
$p_{r,\uparrow}(t)$ is practically indistinguishable for the two cases with
random $c_k$. Even though not shown explicitly here, these two random cases also
coincide for other values of $U$ \cite{SM}. While the two equal cases in Fig.\
\ref{Fig4} differ from the two random ones, this difference is minor in view of
the semi-log plot used. These observations are another central result of
our Letter and clearly show that charge diffusion in the strong-interaction
limit does not depend on the specific initial-state preparation, at least for
the whole class of nonequilibrium states investigated.

{\it Conclusions.} In this Letter, we have investigated the real-time
broadening of charge in the Hubbard chain at high temperatures. First, we have
introduced a class of pure initial states with density profiles where a sharp
peak is located in the middle of the chain and lies on top of a homogeneous
many-particle background. Then, we have calculated the dynamics of these
nonequilibrium states, by using large-scale numerical simulations. Our results
for typical states have unveiled the existence of remarkably clean charge
diffusion in the limit of strong particle-particle interactions, in perfect
agreement with the Kubo formula. We have additionally shown that this diffusive
behavior is stable against varying details of the initial conditions.

{\it Acknowledgments.} We sincerely thank T.\  Prosen and F.\ Heidrich-Meisner
for fruitful discussions. In addition, we gratefully acknowledge the computing
time, granted by the ``JARA-HPC Vergabegremium'' and provided on the ``JARA-HPC
Partition'' part of the supercomputer ``JUQUEEN'' \cite{stephan2015} at
Forschungszentrum J\"ulich.

\setcounter{figure}{0}
\setcounter{equation}{0}
\renewcommand{\thefigure}{S\arabic{figure}}
\renewcommand{\theequation}{S\arabic{equation}}

\section{Supplemental Material}

\subsection{Typicality Approximation}

\subsubsection{Single Projection}

To make our Letter self-contained, we provide details on the
calculation yielding the relation in Eq.\ (\ref{cor}) of the main text. A very
similar calculation can be found in Ref.\ \cite{steinigeweg2016} for the XXZ
spin-$1/2$ chain.

\begin{figure}[t]
\includegraphics[width=0.95\columnwidth]{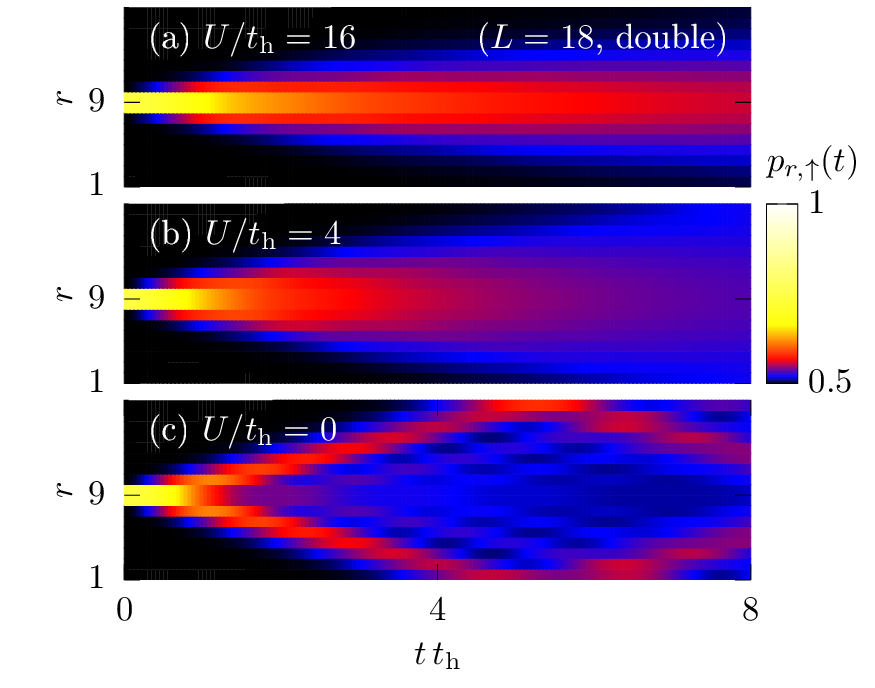}
\caption{(Color online) The simulations in Fig.\ \ref{Fig1} of our Letter
repeated for a random initial state $| \psi_\text{double}(0) \rangle$.}
\label{FigS1}
\end{figure}

Starting with the correlation function
\begin{equation}
C_{r,\uparrow}(t) = 2 \, \langle (n_{L/2,\uparrow} - p_\text{eq.})(n_{r,\uparrow}(t)
- p_\text{eq.}) \rangle + p_\text{eq.} \, , \label{T1}
\end{equation}
multiplying out the two brackets, and using $p_\text{eq.} = 1/2$ as well as
$\langle n_{r,\uparrow}(t) \rangle = p_\text{eq.}$, we get
\begin{equation}
C_{r,\uparrow}(t) = 2 \, \langle n_{L/2,\uparrow} \, n_{r,\uparrow}(t) \rangle =
2 \, \frac{\text{tr}[n_{L/2,\uparrow} \, n_{r,\uparrow}(t)]}{4^L}
\end{equation}
and, due to $n_{L/2,\uparrow}^2 = n_{L/2,\uparrow}$ and a cyclic permutation
in the trace, we obtain
\begin{equation}
C_{r,\uparrow}(t) = 2 \, \frac{\text{tr}[n_{L/2,\uparrow} \, n_{r,\uparrow}(t)
\, n_{L/2,\uparrow}]}{4^L} \, .
\end{equation}
If $| \Phi \rangle$ is a random pure state according to the unitary invariant
Haar measure, this correlation function can be expressed as
\begin{equation}
C_{r,\uparrow}(t) = 2 \, \frac{\langle \Phi | \, n_{L/2, \uparrow} \,
n_{r,\uparrow}(t) \, n_{L/2, \uparrow} \, | \Phi \rangle}{\langle \Phi | \Phi
\rangle} + \epsilon \, ,
\end{equation}
where the negligibly small error $\epsilon \propto 4^{L/2}$ is skipped in the
following for clarity. Because of $n_{L/2,\uparrow}^\dagger = n_{L/2,\uparrow}$,
we can rewrite this expression as
\begin{equation}
C_{r,\uparrow}(t) = 2 \, \frac{\langle n_{L/2,\uparrow} \, \Phi | \,
n_{r,\uparrow}(t) \, | n_{L/2,\uparrow} \, \Phi \rangle} {\langle \Phi | \Phi
\rangle}
\end{equation}
and, applying $n_{r,\uparrow}(t) = e^{\imath H t} \, n_{r,\uparrow} \, e^{-\imath
H t}$ and moving the factor $2$ to the denominator, this expression becomes
\begin{equation}
C_{r,\uparrow}(t) = \frac{\langle e^{-\imath H t} \, n_{L/2,\uparrow} \, \Phi |
\, n_r \, | e^{-\imath H t} \, n_{L/2,\uparrow} \, \Phi \rangle} {\langle \Phi |
\Phi \rangle/2} \, .
\end{equation}
Finally, since $| \psi(0) \rangle = n_{L/2,\uparrow} \, | \Phi \rangle /
\sqrt{\langle \Phi | \Phi \rangle/2}$, we end up with
\begin{equation}
C_{r,\uparrow}(t) = \langle \psi(t) | \, n_{r,\uparrow} \, | \psi(t) \rangle
= p_{r,\uparrow}(t) \, . \label{T2}
\end{equation}
Comparing Eqs.\ (\ref{T1}) and (\ref{T2}) leads to
\begin{equation}
p_{r,\uparrow}(t) -  p_\text{eq.} = 2 \, \langle (n_{L/2,\uparrow} -
p_\text{eq.})(n_{r,\uparrow}(t) - p_\text{eq.}) \rangle \, ,
\end{equation}
i.e., the relation in Eq.\ (\ref{cor}) of the main text.

\begin{figure}[t]
\includegraphics[width=0.95\columnwidth]{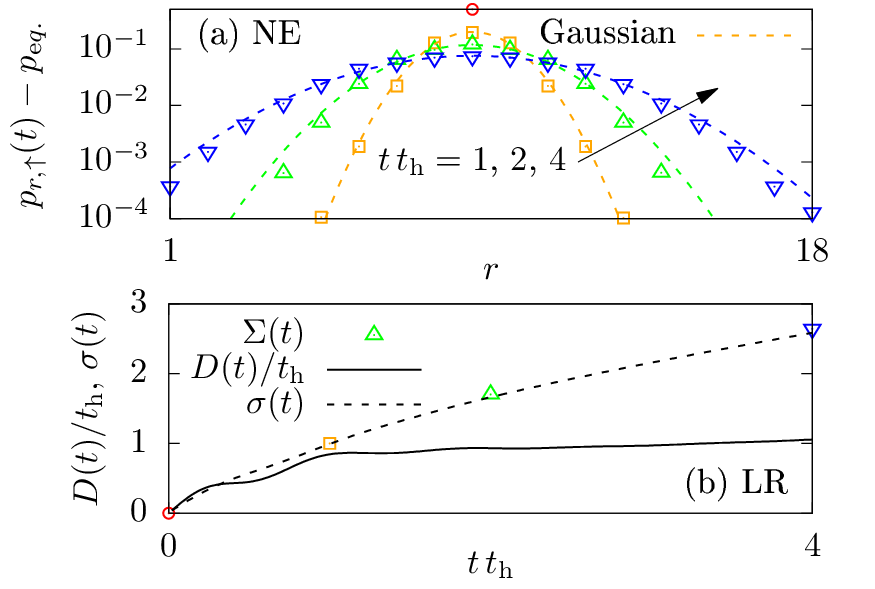}
\caption{(Color online) The same data as Fig.\ \ref{Fig2} of the main text but
for interaction $U/t_\text{h} = 8$ and length $L=18$.}
\label{FigS2}
\end{figure}

\subsubsection{Double Projection}

For the initial state $|\psi_\text{double}(0)\rangle = n_{L/2,\downarrow} \, |
\psi(0) \rangle$, one can simply repeat the steps in Eqs.\ (\ref{T1}) -
(\ref{T2}) to show the relation
\begin{eqnarray}
&& p_{r,\uparrow}(t) -  p_\text{eq.} \nonumber \\
& = & 2 \, \langle (2 \, n_{L/2,\downarrow} \, n_{L/2,\uparrow} - p_\text{eq.})
(n_{r,\uparrow}(t) - p_\text{eq.}) \rangle \, . \label{T3}
\end{eqnarray}
Thus, the resulting relation for $|\psi_\text{double}(0)\rangle$ differs from
the one for $|\psi(0)\rangle$. However, both relations are identical for $U = 0$
(and arbitrary $t$) or for $t = 0$ (and arbitrary $U$). For other values of $U$
and $t$, we can use symmetries of our specific Hamiltonian in Eq.\ (\ref{H}), i.e.,
particle-hole symmetry of each particle species. Due to the symmetry for
spin-down particles, we get
\begin{eqnarray}
&& \langle n_{L/2,\downarrow} \, n_{L/2,\uparrow} \, n_{L/2,\uparrow}(t) \rangle
\nonumber \\
&& = \langle (1- n_{L/2,\downarrow}) \, n_{L/2,\uparrow} \, n_{L/2,\uparrow}(t)
\rangle
\end{eqnarray}
or, equivalently,
\begin{eqnarray}
&& 2 \, \langle n_{L/2,\downarrow} \, n_{L/2,\uparrow} \, n_{L/2,\uparrow}(t)
\rangle \nonumber \\
&& = \langle n_{L/2,\uparrow} \, n_{L/2,\uparrow}(t)
\rangle \, .
\end{eqnarray}
Using this identity, Eq.\ (\ref{T3}) simplifies to Eq.\ (\ref{cor}) and the
relations for $| \psi(0) \rangle$ and $| \psi_\text{double}(0) \rangle$
coincide, at least for the Hubbard chain.

Numerically, we have observed in Fig.\ \ref{Fig4} of our Letter that the
dynamics resulting for the two random initial states $| \psi_\text{double}(0)
\rangle$ and $| \psi(0) \rangle$ are practically the same, at least for the case
of a strong interaction $U/t_\text{h} = 16$. To demonstrate that no difference
is found for other values of $U$ also, we repeat the full calculation in Fig.\
\ref{Fig1} of the main text for $| \psi_\text{double}(0) \rangle$ and depict
the corresponding results in Fig.\ \ref{FigS1}. Apparently, the real-time
broadening does not change for this initial state, in agreement with the
analytical arguments above.

\subsection{Intermediate Interactions}

In Fig.\ \ref{Fig2} of our Letter, we have shown the pronounced Gaussian form of
the density profile $p_{r,\uparrow}(t)$ for different times $t$ but only for a
representative interaction strength $U/t_\text{h} = 16$. To demonstrate that
this form also emerges for other values of $U$, we repeat in Fig.\ \ref{FigS2}
the analysis for the intermediate value $U/t_\text{h} = 8$. As before, the
Gaussian fits in Fig.\ \ref{FigS2} (a) describe the data very well over several
orders of magnitude and the agreement of the resulting standard deviation with
linear response theory in Fig.\ \ref{FigS2} (b) is excellent.

\begin{figure}[t]
\includegraphics[width=0.95\columnwidth]{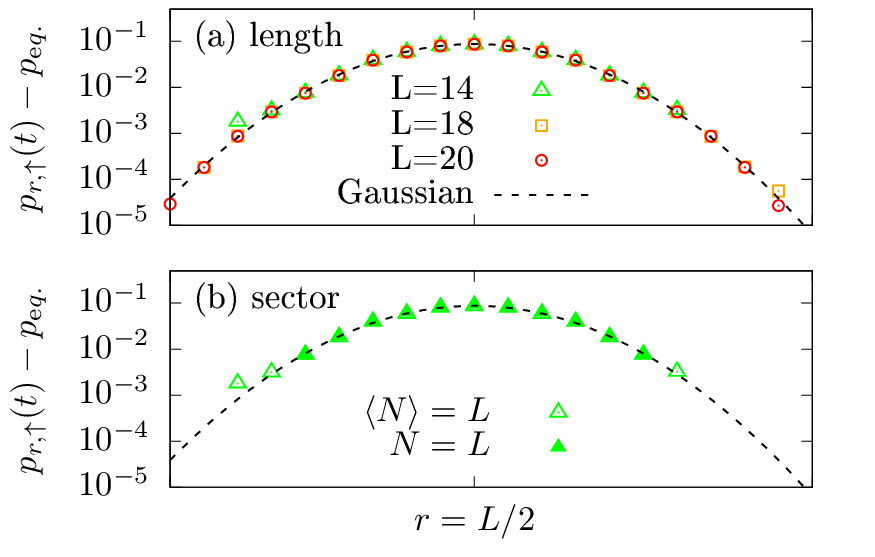}
\caption{(Color online) (a) Density profile $p_{r,\uparrow}(t)$ versus site $r$
for fixed time $t \, t_\text{h} = 4$ and a Hubbard chain with interaction $U/
t_\text{h} = 16$, shown for different lengths $L=14$, $18$, $20$ in a semi-log
plot (symbols) together with a Gaussian fit to $L = 20$ data (curve). (b) The
$L=14$ data in (a) for $\langle N \rangle = L$ in the average, compared to
exactly $N = L$.}
\label{FigS3}
\end{figure}

\subsection{Chain Length and Particle Sector}

All data presented in the main text corresponds to the largest system size
available, i.e., $L=18$ (and $L=20$ for a single set of parameters as well).
In addition, we have performed a careful finite-size analysis to
ensure that these data allow us to draw reliable conclusions on the
thermodynamic limit $L \to \infty$. Thus, as an example, we present in Fig.\
\ref{FigS3} (a) such an analysis for a random initial state $| \psi(0) \rangle$,
interaction strength $U/t_\text{h} = 16$, and the longest time $t \, t_\text{h}
= 4$ considered. Clearly, the density profile $p_{r,\uparrow}(t)$ depicted is
practically the same for lengths $L = 14$, $18$, $20$ and even the outer tails
of this profile do not have a significant finite-size dependence.

Further, it is worth mentioning that we have checked the independence of the
specific ensemble used, i.e., our results do not depend on our choice $\langle
N \rangle = L$. In Fig.\ \ref{FigS3} (b) we demonstrate this independence by a
comparison of the ensembles $\langle N \rangle = L$ in the average and exactly
$N = L$ for the $L=14$ curve in Fig.\ \ref{FigS3} (a), where the finite-size
difference should be strongest. Still, we do not find any significant
difference. However, we should stress that we eliminated trivial finite-size
effects for the $N = L$ case by taking into account that
\begin{equation}
p_\text{eq.} = \frac{L-1}{2L-1} \neq \frac{1}{2}
\end{equation}
for any finite $L$.

\begin{figure}[t]
\includegraphics[width=0.95\columnwidth]{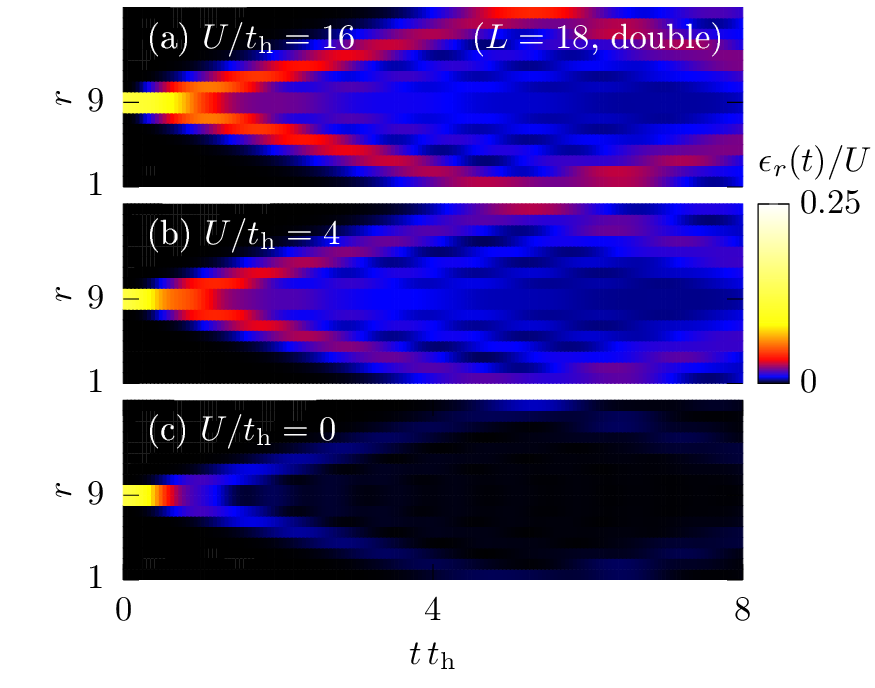}
\caption{(Color online) Time-space density plot of the local interaction energy
$\epsilon_r(t)$ for a random initial state $|\psi_\text{double}(0) \rangle$ in
the one-dimensional Hubbard model with $L=18$ sites and different interaction
strengths: (a) $U/t_\text{h} = 16$, (b) $U/t_\text{h} = 4$, (c) $U/t_\text{h}
= 0$. In contrast to charge, energy flows ballistically in the limit of large
$U$.}
\label{FigS4}
\end{figure}

\subsection{Energy Flow}

In the main text, we have unveiled the emergence of remarkably clean particle
diffusion for sufficiently large interaction strengths $U \gg t_\text{h}$. But
this diffusive behavior should be contrasted against the dynamical behavior of
other quantities for exactly the same initial states. Such a quantity occurs
in the second term of the Hamiltonian (\ref{H}), i.e., the local interaction
energy
\begin{equation}
\epsilon_r(t) = U (n_{r,\downarrow} - \frac{1}{2})(n_{r,\uparrow} -
\frac{1}{2}) \, .
\end{equation}
This quantity becomes the full local energy in the limit of large $U$. Thus,
the flow is expected to be ballistic \cite{karrasch2016-1}, as typical for
integrable systems. In Fig.\ \ref{FigS4} we confirm this expectation for, e.g.,
a random initial state $| \psi_\text{double}(0) \rangle$, where the central
interaction energy $\epsilon_{L/2}(0) = U/4$ takes on the maximum value
possible. Apparently, the jet-like broadening of energy observed for
$U/t_\text{h} =16$ in Fig.\ \ref{FigS4} (a) is very similar to the one of charge
found for $U = 0$ in Fig.\ \ref{FigS1} (c). These findings also illustrate that
our initial states do no enforce diffusive dynamics.

\subsection{Displacement of Particle Species}

For simplicity, our work has focused on the two cases: 1.\ The spin-down
density profiles are $p_{r,\downarrow}(t) = p_\text{eq.}$, i.e., they are at
equilibrium from the very beginning (initial state $| \psi(0) \rangle$). 2.\
These profiles are $p_{r,\downarrow}(t) = p_{r,\uparrow}(t)$, i.e., they are
always identical to the density profiles of the other particle species (initial
state $| \psi_\text{double}(0) \rangle$). This is why we have not shown
results for $p_{r,\downarrow}(t)$ explicitly in the figures.

\begin{figure}[t]
\includegraphics[width=0.95\columnwidth]{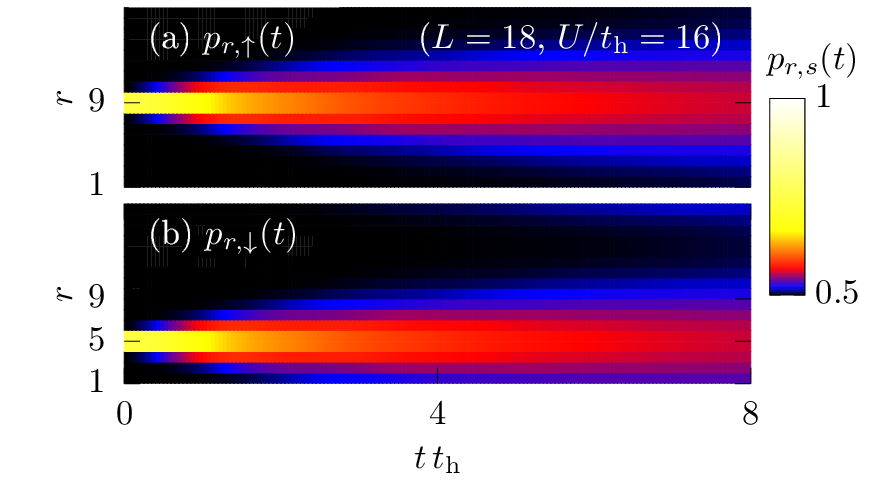}
\caption{(Color online) Time-space density plot of the local occupation numbers
$p_{r,s}(t)$ for both, (a) $s = \uparrow$, (b) $s = \downarrow$ and a Hubbard
chain with $L = 18$ sites and interaction $U/t_\text{h} = 16$, where a random
initial state $| \psi_\text{displaced}(0) \rangle$ is chosen to realize a
displacement of $4$ sites at the beginning. The particle species in (a) and
(b) diffuse independently from each other.}
\label{FigS5}
\end{figure}

Certainly, it is possible to investigate other situations also. For instance,
similar to $| \psi_\text{double}(0) \rangle$, one may study initial states
\begin{equation}
| \psi_\text{displaced}(0) \rangle = n_{L/2-\delta r,\downarrow} \, | \psi(0)
\rangle \, ,
\end{equation}
where $p_{r,\downarrow}(0)$ is not peaked in the middle of the chain and
displaced to the left by $\delta r$ sites. In such a situation,
$p_{r,\downarrow}(t) \neq p_{r,\uparrow}(t)$ and also the dynamical behavior
could change in principle. However, first results indicate that charge dynamics
is qualitatively the same even for this situation. As an example, we show in
Fig.\ \ref{FigS5} results for a random initial state $| \psi_\text{displaced}(0)
\rangle$ with a displacement by $\delta r = 4$ sites, for a Hubbard chain of
length $L = 18$ and with interaction $U/t_\text{h} = 16$. Here, it turns out
that the two particle species spread independently from each other, and no 
significant difference to Fig.\ \ref{Fig1} (a) of the main text is visible.
Once again, this observation can be understood analytically using typicality
and arguments analogous to the ones in the context of Eq.\ (\ref{T3}).

\newpage

\end{document}